\documentclass[pdflatex,sn-basic]{sn-jnl}

\usepackage{graphicx}%
\usepackage{multirow}%
\usepackage{amsmath,amssymb,amsfonts}%
\usepackage{amsthm}%
\usepackage{mathrsfs}%
\usepackage[title]{appendix}%
\usepackage{xcolor}%
\usepackage{textcomp}%
\usepackage{manyfoot}%
\usepackage{booktabs}%
\usepackage{algorithm}%
\usepackage{algorithmicx}%
\usepackage{algpseudocode}%
\usepackage{listings}%
\usepackage{subfigure}%

\theoremstyle{thmstyleone}%
\theoremstyle{thmstyletwo}%

\theoremstyle{thmstylethree}%

\begin{document}

\title[Article Title]{Analysis of Information Digestion Differences\\among Players in Online C2C Markets}


\author*[1]{\fnm{Jun} \sur{Sashihara}}\email{sashihara-jun0116@g.ecc.u-tokyo.ac.jp}
\author[1]{\fnm{Teruaki} \sur{Hayashi}}\email{hayashi@sys.t.u-tokyo.ac.jp}

\affil*[1]{\orgdiv{School of Engineering}, \orgname{The University of Tokyo}, \city{Tokyo}, \country{Japan}}


\abstract{In recent years, the magnitude of consumer-to-consumer (C2C) markets have grown significantly, highlighting the increasing significance of trust between buyers and sellers. However, the specific aspects of product information that facilitate effective communication and trust building in C2C markets remain poorly understood. This study examines the concept of information digestion-the process through which information is accurately understood-and aims to elucidate differences in the information digestion processes of sellers and buyers to clarify the role of product page information. To address this, we conducted two experiments: a questionnaire survey involving 400 subjects and a conjoint analysis with eye-tracking for 15 participants. Initially, we selected eight sample products from four distinct product categories based on the Foote, Cone, and Belding (FCB) grid and compared the product information components considered important by sellers and buyers. Subsequently, we created 12 types of product pages that varied in the combination of four attributes: title, price, product description, and image. Experiment 1 revealed significant differences in the perceived importance of product page components between buyers and sellers. It also demonstrated that product categories and transaction experience influenced the importance assigned to these components, particularly for buyers. Results from Experiment 2 showed that buyers prioritize product descriptions and identified two distinct buyer groups based on their information digestion patterns: one that carefully reads and digests information and another that processes information more rapidly. These findings enhance our understanding of trust-building mechanisms in online C2C markets and provide practical insights for platform designers and market participants.}




\maketitle

\section{Introduction}\label{intro}
The rise of digital technology in contemporary society has led to the significant expansion of consumer-to-consumer (C2C) markets (\cite{emarketer2018}), profoundly influencing buying habits (\cite{yuan2021review}). Mercari, a prominent C2C marketplace app in Japan, exemplifies this trend by offering products across more than 1,000 categories ranging from men's polo shirts to laptops. These platforms enable individuals to buy and sell a wide variety of goods. In C2C online marketplaces, such as Mercari, product information is shared by sellers on dedicated pages, which buyers then use to make purchasing decisions. Consequently, the comprehensiveness and quality of this information play a crucial role in understanding products and purchasing choices (\cite{lewis2011asymmetric}). However, issues such as information overload and discrepancies in communication between buyers and sellers can undermine trust, reduce market efficiency, and impede transactions (\cite{leonard2021trust}). For example, irrelevant information or tags on product pages may prevent buyers from accessing useful information. By contrast, the absence of critical product information can prevent buyers from accurately assessing an item. Therefore, to foster effective communication and sustainable growth in online C2C markets, it is vital to understand the disparities between buyers and sellers in terms of \textit{information digestion} (\cite{koike2020digestion}). This concept focuses on how well the recipient correctly understands information after receiving it.

This study examines how buyers and sellers digest information differently to improve communication efficiency in online C2C marketplaces. It addresses three research questions. (1) Are there disparities in the way buyers and sellers digest information? (2) How do factors such as the type of product and transaction history influence the processing of information digestion? (3) What are the distinct features of buyer's information digestion? Specifically, our goal is to conduct a quantitative analysis of how various components on product pages affect the information digestion of both parties, and to explore how these differences impact the behavior and choices of market participants. 

\begin{figure}[h]
\centering
\includegraphics[width=0.7\textwidth]{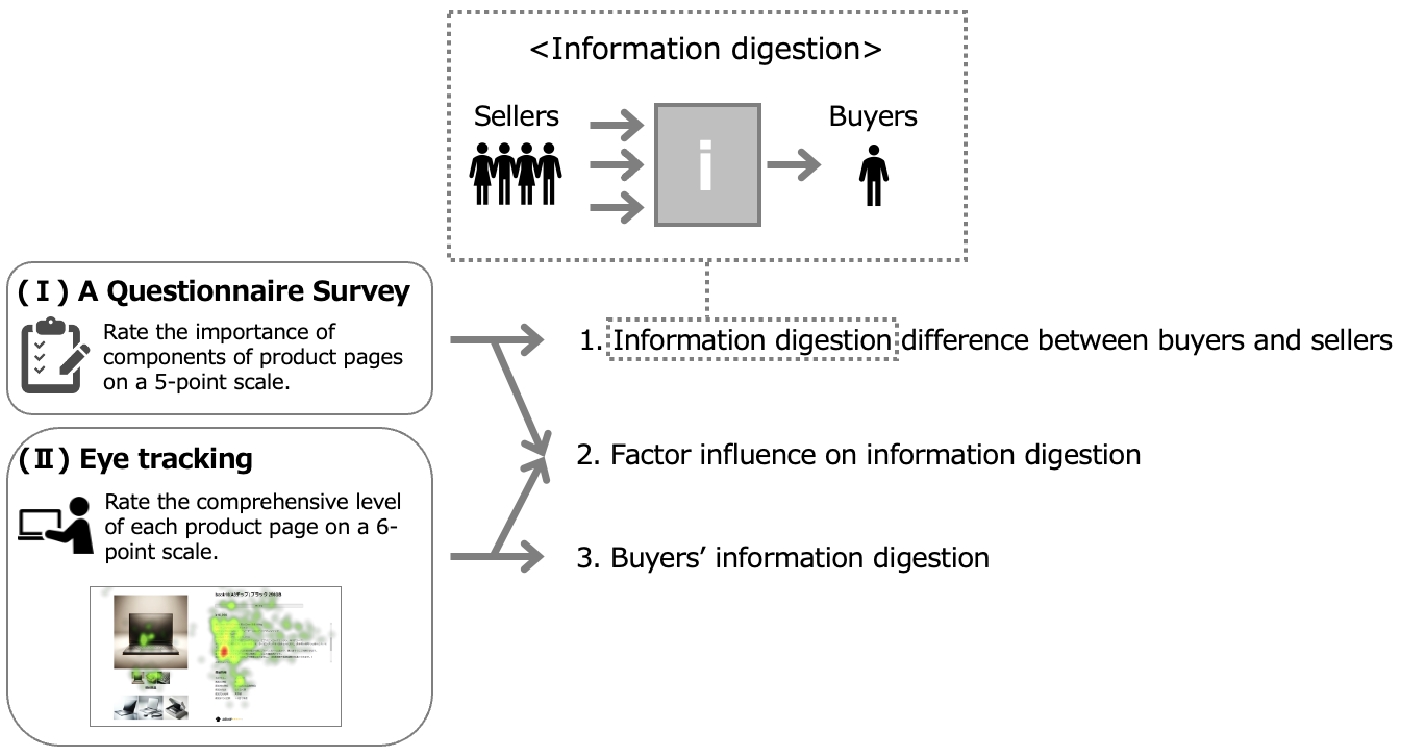}
\caption{Overview of three research questions and two experiments in this study.}\label{fig:research_overview}
\end{figure}

To address these issues, two experiments were conducted (Figure \ref{fig:research_overview}). First, a questionnaire survey and statistical analysis were performed to identify the differences in the information that buyers and sellers consider important. The survey used the Foote, Cone and Belding (FCB) grid (\cite{vaughn1980advertising}), a framework for categorizing products based on how consumers make purchasing decisions, to comprehensively cover various product categories. Second, a page view and eye-tracking experiment was conducted with buyers to explore the details of their information digestion. In this experiment, a conjoint analysis was carried out by varying the content of product page elements to identify the information that influences buyers' understanding. 

This study contributes to the development of more user-friendly C2C market platforms. By analyzing the key components of product pages and their impact on information digestion, this study offers insights that could lead to the enhancement of product pages and the fostering trust in online C2C markets.

The remainder of this paper is organized as follows: Related studies are discussed in Section \ref{related work}. Section \ref{experiments} outlines the settings of Experiment 1, which involves a questionnaire survey to identify differences between buyers and sellers, and Experiment 2, a practical experiment with buyers using an eye-tracking device. Based on these experiments, the results are presented and discussed in Sections \ref{results} and \ref{discussion}. Finally, conclusions and future directions are summarized in Section \ref{conclusion}.

\section{Related Works and Our Approach}
\label{related work}
\subsection{Information Overload and Digestion}
\label{information overload and digestion}
The volume of information has increased rapidly since the beginning of the 21st century. For example, Bawden \textit{et al.} noted that more information was created in the last 30 years of the 20th century than in the preceding 5,000 years (\cite{bawden2009dark}). In 2012, approximately 2.5 exabytes of data were generated daily, and this rate of data generation doubles every three years. Currently, the data transmitted every second through the Internet exceeds the total data available on the Internet 20 years ago (\cite{mcafee2012big}). This explosion of information has led to information overload, a phenomenon where excessive information, even if relevant and valuable, becomes problematic rather than beneficial (\cite{bawden2009dark}).

Although information overload has existed since the earliest records of information, its nature and causes have evolved over time (\cite{bawden2020information}). It was initially recognized as an academic issue with the advent of publishing technology but has since escalated into a broader social problem (\cite{bawden1999perspectives}). The growing emphasis on big data has further intensified this challenge (\cite{Merendino2018}). Information overload has been shown to reduce happiness and satisfaction, contribute to feelings of self-loss, and create confusion in decision-making (\cite{lee2004effect, kominiarczuk2014turn}). Additionally, it has exacerbated social problems such as proliferation of fake news and clickbait articles (\cite{lazer2018science}).

To mitigate information overload, \cite{koike2020digestion} introduced the concept of \textit{information digestion}, which focuses on how effectively a recipient understands information after receiving it. They evaluated the efficiency of information digestion through a hierarchical factor analysis based on a questionnaire survey covering four media types: online news articles, online advertisements, online shopping platforms, and academic articles or reports. 

\subsection{Consumer-to-Consumer Market}
\label{c2cmarket}
Online transactions are on the rise, with online retail sales projected to account for 20\%  of the market by 2021 (\cite{emarketer2018}). Among these, C2C e-commerce has experienced particularly rapid growth (\cite{wei2019trust}). A notable examples is eBay, a platform designed to facilitate transactions between individual consumers and support the buying and selling of goods (\cite{makelainen2006b2c}). In Japan, Mercari holds the largest market share, successfully increasing its sales by using its user-friendly interface and extensive product offerings. The platform's product pages include seven components designed to enhance usability. In addition, online communities such as Facebook and WeChat are increasingly supporting C2C transactions, with social media platforms improving economic efficiency. For instance, Instagram has developed market features, including live-streaming product sales and creating trading communities (\cite{sukrat2016evolution}).

C2C transactions, typically conducted between individuals, are facilitated through online marketplaces and auction sites. One major advantages of C2C transactions is their low cost. However, building trust between buyers and sellers remains a significant challenge. In contrast, business-to-business (B2B) transactions occur between companies, where trust is more easily built. B2B transactions are often repeated, fostering the establishment of long-term relationships. 

\subsection{Our Approach}
The problem of information overload and the methods for facilitating C2C transactions have  traditionally developed independently. However, trust and sales are closely interconnected, with information overload and the trustworthiness of the information in the C2C market directly influencing transaction outcomes. For example, a good reputation enhances trustworthiness and significantly affects buyers' purchasing decisions (\cite{leonard2021trust}). Joo showed that in auction-style transactions, buyers are more willing to spend on expensive products when they trust the sellers (\cite{joo2015roles}). Similarly, Bao \textit{et al.} found through a survey of online shopping in China that seller trustworthiness positively influences buyers' repurchase intentions (\cite{bao2016repurchase}). This connection between trustworthiness and purchasing behavior is particularly pronounced for buyers with limited online shopping experience (\cite{leonard2021trust}). Trustworthiness, an important factor in C2C transactions, is shaped by various elements, including perceived  website quality and third-party endorsements (\cite{yoon2015influencing, jones2008trust}). Moreover, the quality of product pages and the clarity of the information provided affect not only trustworthiness but also cost efficiency and time management.

The absence of established concepts regarding information digestion in C2C transactions gives rise to two primary issues: information selection and interpretation. Sellers use various methods to capture buyers' attention and convey product information, but some methods-such as unnecessary tagging or misleading product descriptions-can hinder buyers' ability to effectively select relevant information. Even when essential information is included on a product page, misinterpreting the seller's intent can lead to miscommunication, diminished trust, and reduced likelihood of repeat purchases. This study considers the differences in the viewpoints between sellers and buyers, the components of product pages, and the missing categories in information digestion, a framework proposed by \cite{koike2020digestion}. By analyzing and categorizing the processes of information digestion for buyers and sellers, it may be possible to provide tailored, optimized information that enhances communication and trust in C2C marketplaces.

To address these problems, this study applies the concept of information digestion, focusing on understanding of information rather than merely its provision, to identify the elements that influence buyers' comprehension. First, we examine whether information communication gaps exist between buyers and sellers in the C2C market. We conducted a questionnaire survey (Experiment 1) where participants assumed the roles of buyers and sellers. By analyzing responses based on these roles, we explored the differences in purchasing and selling behaviors. The core approach involves deconstructing the product page into specific components, such as the title and description, each conveying distinct types of information. Few studies have analyzed C2C product pages with this item-centric perspective. Furthermore, we included eight product categories in our analysis, as we hypothesize that the information digestion process may vary by product category, similar to Koike's observation that information digestion methods differ across information media (\cite{koike2020digestion}).

Next, we investigate buyers' information digestion through product page evaluations and eye-tracking experiments (Experiment 2). This experiment aimed to simulate an actual purchasing scenario more realistically than the questionnaire-based evaluations in Experiment 1. Participants evaluated the comprehensibility of product pages while their eye movements were simultaneously tracked. We employed conjoint analysis to evaluate the significance of product page components. Eye-tracking was used to identify which components participants physically focused on. While many marketing studies have used eye-tracking technology to optimize website designs and sales strategies (\cite{SoussanDjamasbi2010}), there is limited research applying the technology to the consumer-to-consumer (C2C) market.

\section{The Detailed Explanation of Experiments 1 and 2}
\label{experiments}

In this study, we examined product pages in the C2C market to identify components that promote high information digestion efficiency, focusing on basic page elements such as title, description, and image. Through two experiments, we tested three hypotheses related to the research questions outlined in Section \ref{intro}.
\begin{itemize}
\item Experiment 1: Analyzed differences in information digestion between buyers and sellers through a questionnaire survey, focusing on the product page components each group prioritizes. 
\item Experiment 2: Investigated buyers' information digestion using product page evaluations and eye-tracking experiments to understand how buyers process and prioritize product page components.
\end{itemize}

\subsection{Experiment 1: A Questionnaire Survey}
\subsubsection{Objective and Methodology}
The objective of Experiment 1 is to determine whether differences exist in information communication between sellers and buyers in the C2C market. We posit that the information conveyed by a product page is the cumulative contribution of its components: \textit{ title, price, description, product information, seller information, image}, and \textit{similar products}. To explore these differences, we conducted a questionnaire survey to identify which components buyers prioritize when purchasing and which components sellers prioritize when creating product pages. The survey question was, ``Please rate the importance of XX (components, such as title and description) on a 5-point scale to understand/create the product page in the YY category”.

\subsubsection{Participants and Preparation}
The participants in this study were individuals in their 20s who browsed the C2C product pages at least once a month. This age group was chosen because they frequently engage in online purchases and reflect current market trends. The participant ratio was set at 6:4 based on whether they had prior transaction experience. A preliminary survey was conducted with 20,000 participants through the research company PLUG-Inc., from which 400 participants (200 men and 200 women) meeting the inclusion criteria were selected. The 400 participants were evenly divided into two groups of 200 each: one group acted as buyers, and the other as sellers. This survey was conducted after receiving approval from the Ethics Committee of the Graduate School of Engineering at the University of Tokyo (examination number: 23-122; approval number: KE23-128).

\subsubsection{Experiment Design}
Assuming that the information digestion process differs depending on the product category, we set eight categories based on the FCB grid (\cite{vaughn1980advertising}). The FCB grid classifies products into a $2\times2$ matrix using two axes: involvement (high or low) and product type (thinking or feeling). It is commonly used to analyze consumer buying behavior and design advertising strategies. Each product group in the FCB grid is characterized as follows. Products with high-involvement thinking are usually purchased after collecting detailed information and evaluating functionality. Products with high-involvement feeling are influenced more by emotional factors such as brand image and design. Products with low-involvement thinking products are usually purchased out of habit, with minimal concern for brand loyalty. Products with low-involvement feeling are typically impulse purchases driven by the desire for personal satisfaction.

We used this model to include a wide range of product categories in our questionnaire survey. Two categories were selected from each of the four quadrants of the $2\times2$ matrix. Consequently, eight categories were selected (Table \ref{tab:FCB grid}): \textit{Smartphones} and \textit{Laptops} (high-involvement thinking), \textit{Men's Polo Shirts} and \textit{Women's Polo Shirts} (high-involvement feeling), \textit{Daily Necessities} and \textit{Laundry Detergents} (low-involvement thinking), and \textit{Wine} and \textit{Water} (low-involvement feeling).

\begin{table}[h]
\centering
\caption{Eight sample product categories in four types based on the FCB grid}
\label{tab:FCB grid}
\begin{tabular}{l|ll}
\toprule
Types  & Product 1 & Product 2 \\
\midrule\midrule
High-Involvement Thinking & Smartphone & Laptops\\
High-Involvement Feeling & Men’s Polo Shirts & Women’s Polo Shirts\\
Low-Involvement Thinking & Daily Necessities & Laundry Detergents \\
Low-Involvement Feeling & Wine & Water\\
\botrule
\end{tabular}
\end{table}

\subsubsection{Data Collection Procedure}
A screening survey was conducted prior to the main survey to ensure data reliability and precision. Participants were selected based on their browsing habits and experiences with C2C product pages. It is important to note that the survey was conducted online.

\begin{figure}[h]
\centering
\includegraphics[width=\textwidth]{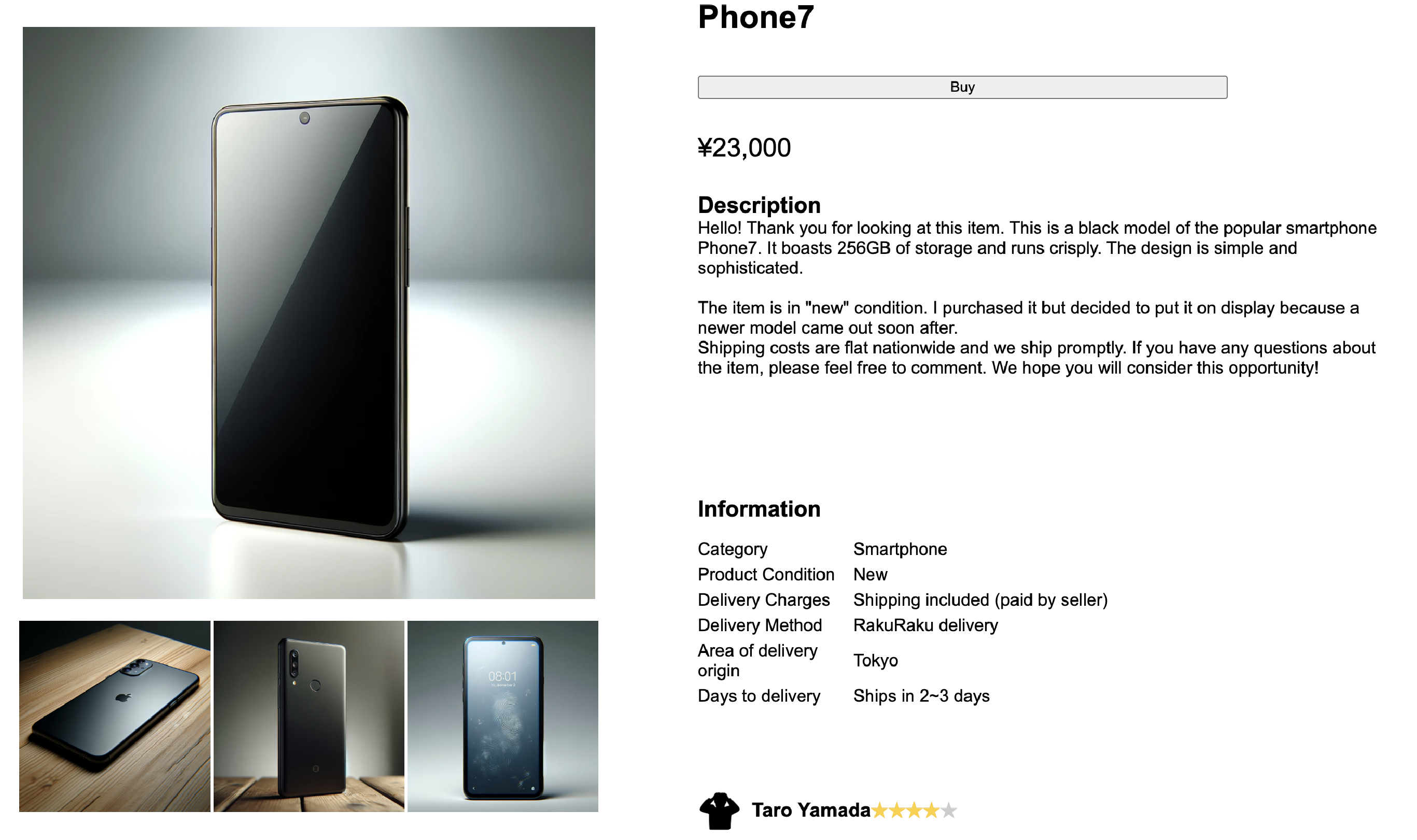}
\caption{An example of a product page presented to participants. This figure is translated into English, but participants were showed in Japanese.}\label{fig:smartphone}
\end{figure}

The following procedure was conducted with the participants:
\begin{enumerate}
\item The experiment was explained, and consent for participation was obtained.
\item Participants were asked questions to evaluate the importance of product page components for the first category.
\item Participants were then asked similar questions for the second category.
\end{enumerate}
After being shown an example product page (Figure \ref{fig:smartphone}), participants rated the importance of the seven components of the product page on a 5-point scale (1 = not at all important, 2 = not very important, 3 = neither important nor unimportant, 4 = somewhat important, and 5 = very important) for understanding and creating a product page. Each participant responded to questions about product pages for two categories selected from eight options: smartphones, laptops, men's polo shirts, women's polo shirts, daily necessities, laundry detergents, wine, and water. For example, if the question was, ``Please rate the importance of \textit{title} on a 5-point scale to understand the product page in \textit{smartphones category}.", a sample response might be ``4." The order of presentation for each category was randomized to avoid bias.

\subsubsection{Analysis}
We conducted statistical analysis to examine the survey data distributions for buyers and sellers. First, we employed the Mann-Whitney U test, a nonparametric method used to determine whether significant differences exist between the distributions of two independent groups. Since the Shapiro-Wilk test confirmed that all component distributions were nonparametric, the Mann-Whitney U test was used to analyze differences in information digestion between buyers and sellers.

Next, we applied the Aligned Rank Transform (ART) test, a nonparametric approach analogous to a two-way ANOVA. This analysis revealed that both product category and transaction experience significantly affect information digestion processes of the buyers.

\subsection{Experiment 2: Conjoint Analysis and Eye-Tracking}
\subsubsection{Objective and Methodology}
We conducted Experiment 2 to investigate buyers' information digestion processes. Participants evaluated product pages while their eye movements were simultaneously tracked. As in Experiment 1, we hypothesized that the information digestion process would differ by product category and, therefore, used the same eight categories based on the FCB grid.  Each participant rated four of the eight categories. For each category, we created 12 product pages with varying information content across components and performed a conjoint analysis. Participants rated each product page comprehensively on a 6-point scale. 

Conjoint analysis is a widely recognized method for analyzing consumer decision-making processes (\cite{tripathi2010empirical,rao2010conjoint}). For instance, Sung and Chung used conjoint analysis to examine online shopping behavior, incorporating six attributes, such as price and the number of reviews (\cite{sung2023factors}). Moreover, combining conjoint analysis with eye-tracking can provide deeper insights into consumer preferences (\cite{behe2014incorporating, meissner2010eye, meissner2016eye}). Eye-tracking is a valuable tool in marketing science, as humans naturally focus on visual stimuli. For example, Meissner \textit{et al.} showed that repeated tasks using eye-tracking improved participants' understanding of key points of attention, enhancing the reliability of their choices and reducing processing time (\cite{meissner2016eye}). Similarly, Djamasbi \textit{ et al.} found that Generation Y preferred webpages featuring large images (\cite{SoussanDjamasbi2010}). Additionally, Xu and Liu reported the impact of visual processing on purchase decisions using machine learning and EEG data, offering a novel perspective on consumer behavior. (\cite{xu2024decoding}).

\subsubsection{Participants and Preparation}
The participants consisted of 16 students in their 20s who browsed the C2C product pages at least once per month, reflecting current consumer trends. All participants took part in the on-site experiments, and each received a 2,000 Japanese yen Amazon gift card upon completing the experiment. Each session lasted approximately 30 min. During the analysis, data from one participant were excluded as an outlier because the participant gave identical ratings to all product pages. 

Before the experiment, participants were informed of its purpose and provided signed consent forms. They were seated in a chair at a desk with a laptop (Panasonic Let's note CF-SR) that had a screen resolution of $1920\times1080$. Eye movements were tracked using Tobii Pro Spark eye tracker, which was attached to the bottom of the laptop screen. The Tobii Pro Lab software was used to record data, with the top-left corner of the display serving as the origin point (0,0) (\cite{TobiiProLab, TobiiProLabManual}). Prior to the experiment, participants underwent a calibration process to estimate the geometric characteristics of their eyes. During this process, participants were instructed to keep their heads as still as possible to ensure accurate calibration and reliable eye-tracking data. 

\begin{figure}[h]
\centering
\subfigure[]{
\includegraphics[width=0.35\textwidth]{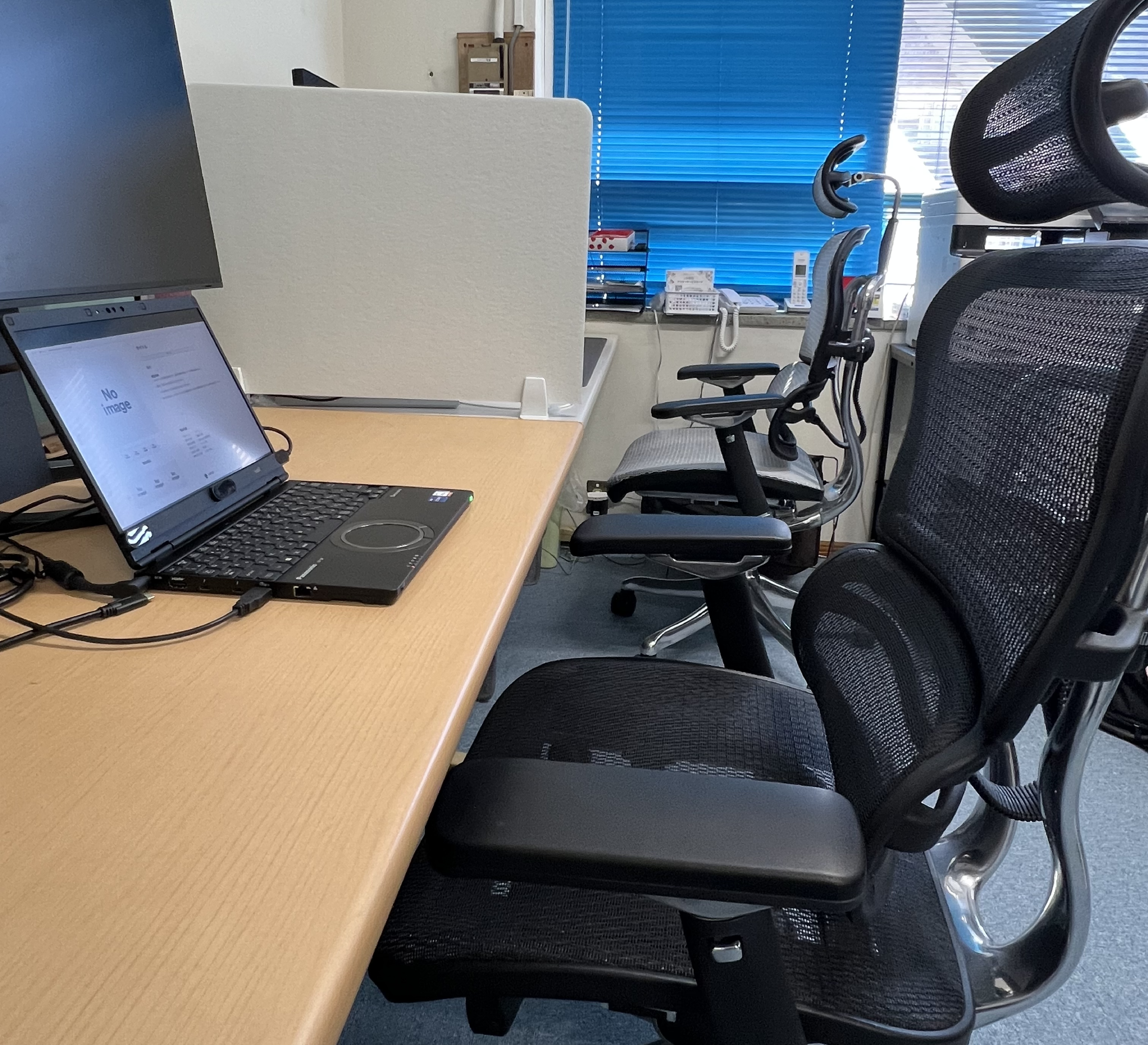}
}~
\subfigure[]{
\includegraphics[width=0.65\textwidth]{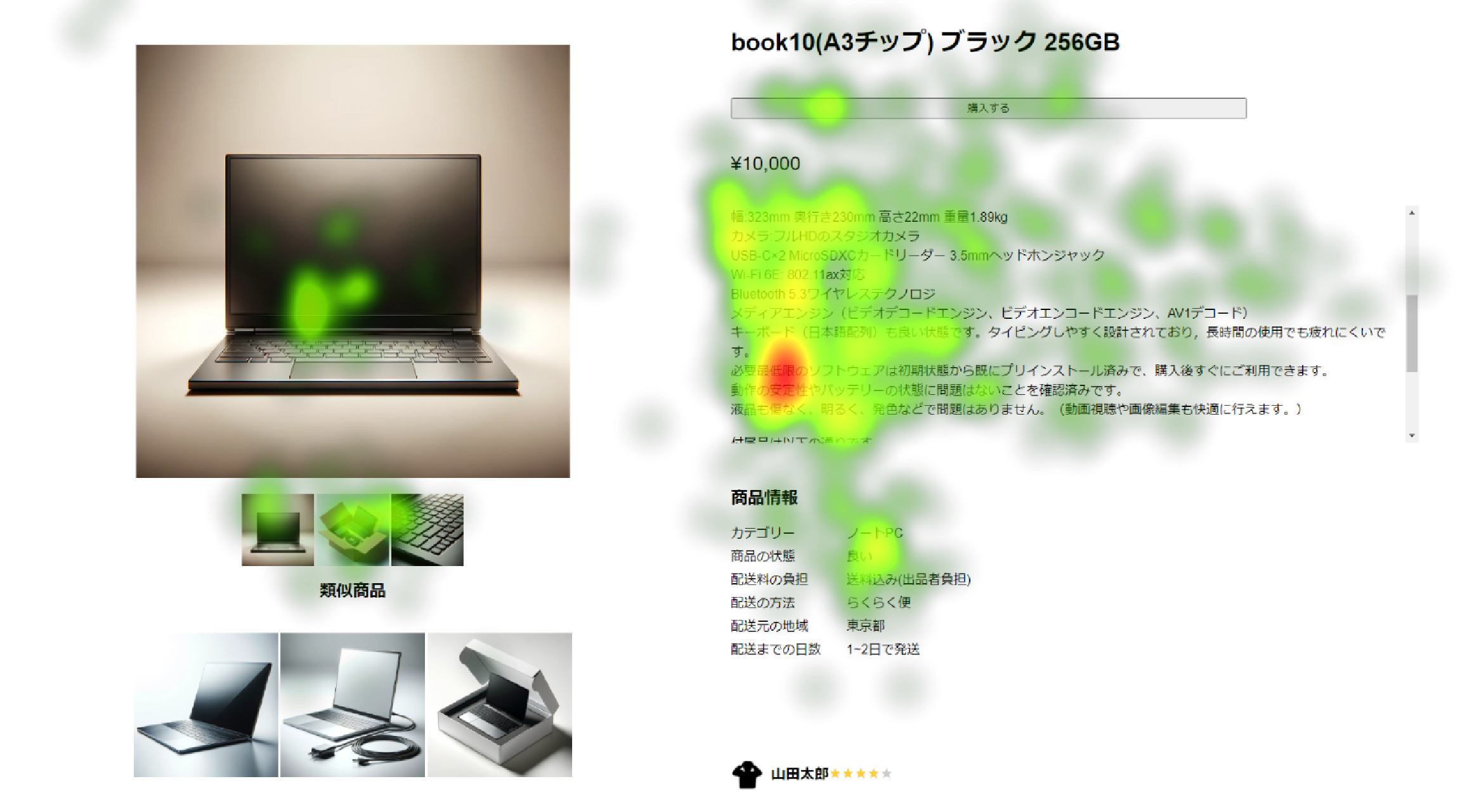}
}
\caption{(a) Experimental environment for Experiment 2, and (b) an example result obtained by the eye tracker. In this example, participant gazed title, description, product information, and image on product page. Description is viewed most.}\label{fig:experiment_condition}
\end{figure}

\subsubsection{Experiment Design}
We assumed that a product page is composed of seven elements: title, price, description, product information, seller information, image, and similar products. As in Experiment 1, we hypothesized that the information digestion process varied between different product categories. We used the same eight sample products from the four types based on the FCB grid as in Experiment 1. To empirically identify the important page components for buyers' information digestion, we prepared 12 pages for each category. Each page contained different sets of information across four components: title, price, description, and image. We chose to vary the content of four of the seven components because changing all components would require an impractically large number of pages. Title, description, price, and image were selected because the sellers can freely determine and customized their content. Furthermore, preliminary experiments suggested that these four components significantly impact buyers' information digestion. In the preliminary experiment, four participants viewed product pages from eight categories for at least 30 s each. After browsing, participants identified the components that made the page most understandable for each category. The results showed that description was the most important component, followed by title, price, and image.

Different levels were set for each component. We assigned two levels to title, price, and image, and three levels to description. The description was divided into three levels because preliminary experiments revealed its greater potential to influence buyers' information digestion. The two levels for title were short and long, for price were low and high, and for image were one and three. The three levels of description were short, medium, and long (Table \ref{tab:conjoint_analysis_setting}). The specific character counts and price values for each level were determined by analyzing actual product data. For each of the eight categories, we randomly selected 100 product data from Mercari. The first and third quartiles of character counts were used for title and price values. For description, we used the first, mean, and third quartiles of character counts. For example, in the smartphone category, the level 0 title is ``Phone 7." and the level 1 title is ``Phone7 black 256GB unlocked." The content of the components varied by category to more closely reflect the real-world data. 

\begin{table}[h]
\centering
\caption{Content levels of four components on product pages for Experiment 2.}
\label{tab:conjoint_analysis_setting}
\begin{tabular}{ccccc}
\toprule
Level & Title & Price & Description & Image \\
\midrule\midrule
0 & short & low & short & one image\\
1 & long & high & middle & three images\\
2 & - & - & long & - \\
\botrule
\end{tabular}
\end{table}

Considering the combinations of levels for the four components, we adopted a fractional factorial design. Combining all levels yielded 24 ($= 2\times2\times3\times2$) hypothetical product pages for each category. Since participants were tasked with browsing pages from four categories, this would require them to evaluate 96 (= 24 pages $\times$ 4 categories) pages. However, evaluating such a large number of pages could be too overly burdensome for the participants. To address this, we used a fractional factorial design to reduce the number of pages evaluated to 48 (= 24 $\times \frac{1}{2} \times$ 4 categories). Table \ref{tab:fractional_factorial_design} shows the combinations of component levels for each product page. The product pages were designed using HTML, CSS, and JavaScript and modeled after read-world product pages. Text sizes and fonts were standardized across all pages for consistency. To avoid using actual products, we used DALL-E3 \footnote{https://openai.com/dall-e-3} to generate product images. DALL-E3, an AI-based image generation model developed by OpenAI, creates images from natural language descriptions, ensuring the authenticity of the experiments while avoiding real product associations.

\begin{table}[h]
\centering
\caption{Combinations of attribute levels for 12 product pages using a fractional factorial design}
\label{tab:fractional_factorial_design}
\begin{tabular}{c|cccc}
\toprule
Page Number & Title & Price & Description & Image \\
\midrule\midrule
1 & 0 & 0 & 0 & 0 \\
2 & 0 & 0 & 2 & 0 \\
3 & 0 & 1 & 0 & 1 \\
4 & 0 & 1 & 2 & 1 \\
5 & 0 & 0 & 1 & 1 \\
6 & 0 & 1 & 1 & 0 \\
7 & 1 & 0 & 0 & 1 \\
8 & 1 & 0 & 2 & 1 \\
9 & 1 & 1 & 0 & 0 \\
10 & 1 & 1 & 2 & 0 \\
11 & 1 & 0 & 1 & 0 \\
12 & 1 & 1 & 1 & 1 \\
\botrule
\end{tabular}
\end{table}

\subsubsection{Data Collection Procedure}
Data were collected in our laboratory in December 2023. After a demonstration using a test screen, participants browsed each product page for at least 20 s while an eye tracker recorded their eye movements. Following each browsing session, participants rated the comprehensibility of the product page on a 6-point scale (1 = most difficult to understand, 6 = easiest to understand) and reported their ratings orally. To minimize head and eye movements unrelated to browsing, the researcher controlled the transitions between pages. Additionally, the order of the four categories and the 12 pages within each category was randomized to mitigate order bias.

\subsubsection{Analysis}
Participants' 6-point ratings of page comprehensibility were analyzed using conjoint analysis. In this conjoint analysis, the importance of each component was determined by assigning dummy variables to the levels of each component in the experiment and treating these variables as explanatory variables. Since the experiment included two levels for title, price, and image, and three levels for description, two dummy variables were assigned to title, price, and image, and three dummy variables were assigned to description. 

Using the participants' evaluation data as the response variable, we performed Quantification Theory Type 1 to calculate the part-worth utility for each level of the four components. The importance of each component in the information digestion process was then quantified by calculating the range of part-worth utilities for each component. This range was calculated by subtracting the smallest part-worth utility from the largest value within each component $A$, as shown in Equation \ref{range}.
\begin{equation}\label{range}
range(A) = max(A) - min(A)   
\end{equation}
These Quantification Theory Type 1 calculations were performed using 12 data points for each combination of participants and categories (e.g., Participant 1's evaluation of laptops). For example, if the part-worth utilities for the three levels of description ($D1, D2, D3$) for Participant 1's laptops evaluation were 1.0, 2.0, and 3.0, respectively, the range for the description was calculated as $D3 - D1 = 3 - 1 = 2$. Therefore, 60 sets of importance values (15 participants $\times$ 4 categories) were obtained. We analyzed data from 15 participants; however, the small sample size limits the generalizability of the results. Nevertheless, having each participant evaluate four categories helped mitigate these limitations. Using the importance values, we identified the components that buyers prioritize. Furthermore, we classified the different information digestion processes through clustering.

\begin{figure}[h]
\centering
\includegraphics[width=\textwidth]{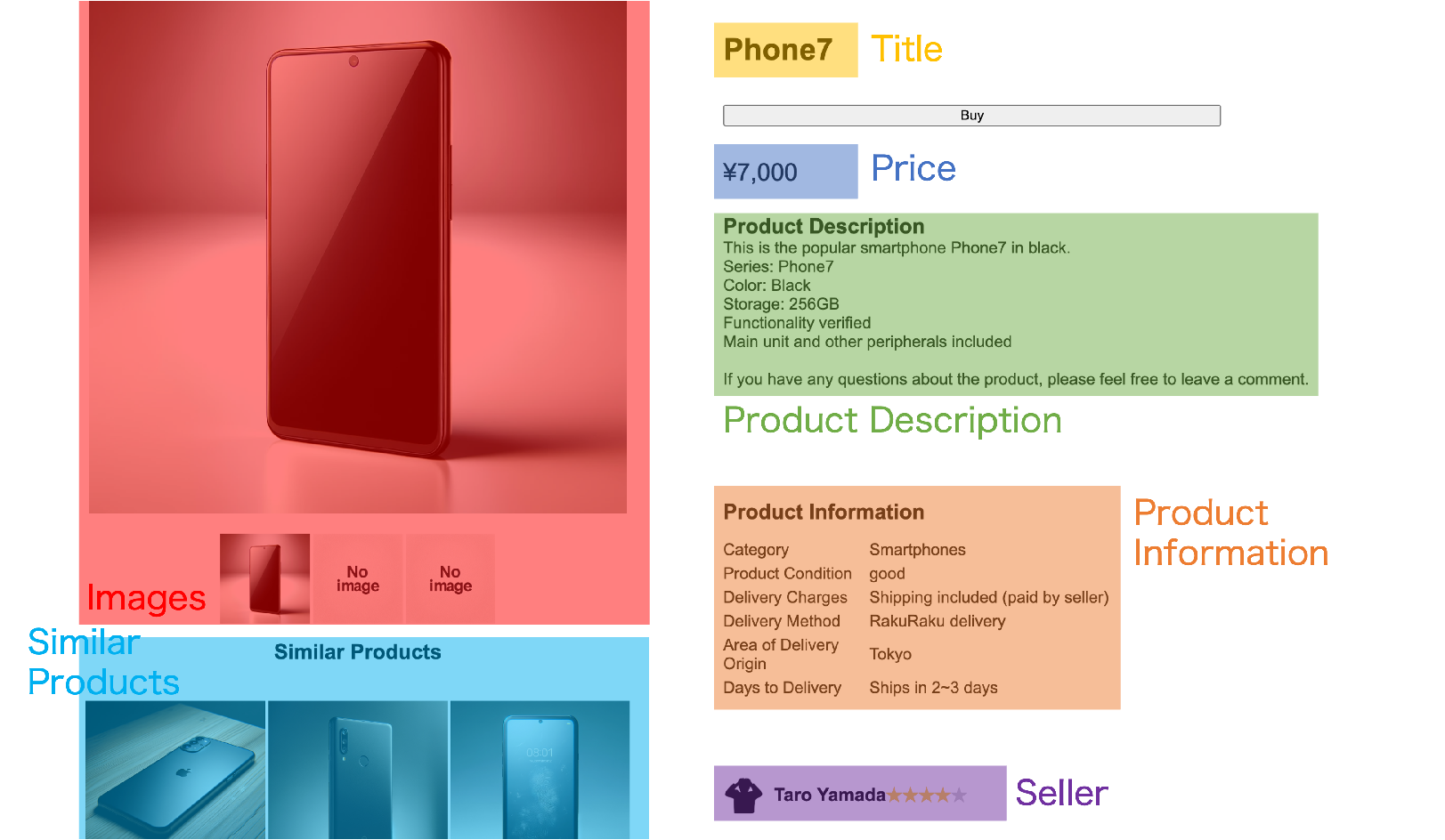}
\caption{Setting AOIs in this study}\label{fig:pre_experiment_AOI_setting}
\end{figure}

In the eye-tracking analysis, we used the following four indicators, as described by Behe \textit{et al.} (\cite{behe2014incorporating}): time to first fixation (TFF), first fixation duration (FFD), total visit duration (TVD) within the area of interest (AOI), and fixation count (FC) within the AOI. The TFF measures the time taken to fixate on an AOI after the screen is presented. The FFD indicates the time a participant spends during the initial fixation on an AOI. The TVD is the total time spent looking at an AOI, and FC is the number of times a participant fixates on an AOI. The AOI refers to specific areas or elements that attract participants' attention. In the eye-tracking experiment, participants' attention (eye movements) was classified into saccades and fixations. Saccades are rapid eye movements used to gather information, while fixations occur when the eye focus on a specific area for a period of time. In Experiment 2, we designated seven page components as AOIs (Figure \ref{fig:pre_experiment_AOI_setting}) and defined saccades and fixations using Tobii's system based on the angular velocity of eye movements. We interpreted buyers' eye movements using these four indicators. In addition, by comparing the correlation coefficients between the results of the conjoint analysis and the eye-tracking indicators across clusters, we further explored the features of information digestion.

\section{The results of Experiments 1 and 2}
\label{results}
\subsection{Experiment 1}
There were two main results from Experiment 1: First, the importance of page components differs for buyers and sellers. We conducted a Mann-Whitney U test on the 5-point importance evaluation data for the page components (Table \ref{tab:mannwhitneyu}). The results show that seven page components have p-values of less than 0.05, indicating a significant difference in the importance of page components between buyers and sellers. Although the medians were the same, a significant result in the Mann-Whitney U test suggests differences in the spread (variability) or shape (such as skewness) of the distributions. As shown in Table \ref{tab:mannwhitneyu}, the standard deviation of the buyer data is greater than that of the seller data. 

Second, there was variation in the components that buyers considered important. The Aligned Rank Transform (ART) test indicated that two factors, product category and transaction experience, influenced the importance of page components. The category effect was significant for four components: price, description, product information, and image ($p < 0.01$). Furthermore, the influence of purchasing experience was significant for five components: price, description, product information, seller information, and image. However, no significant effect of product category on any component was observed for sellers. The main effect of the sales experience was significant only for title.

\begin{table}[h]
\caption{Mann-Whitney U test for each item on product page between buyers and sellers.}\label{tab:mannwhitneyu}
\begin{tabular*}{\textwidth}{@{\extracolsep\fill}lcccccccc}
\toprule%
& \multicolumn{3}{@{}c@{}}{Buyer ($n=200$)} & \multicolumn{3}{@{}c@{}}{Seller ($n=200$)} & P-Value \\ \cmidrule{2-4}\cmidrule{5-7}%
 & Mean & Median & SD & Mean & Median & SD & \\
\midrule\midrule
title & 3.72 & 4 & 1.15 & 4.15 & 4 & 0.95 & **\\
price & 4.49 & 5 & 0.85 & 4.64 & 5 & 0.69 & **\\
product description & 4.26 & 5 & 0.92 & 4.37 & 5 & 0.80 & *\\
product information & 4.13 & 4 & 1.02 & 4.24 & 4 & 0.92 & *\\
seller information & 3.38 & 4 & 1.27 & 3.20 & 3 & 1.21 & **\\
image & 4.00 & 4 & 1.10 & 4.34 & 5 & 0.89 & **\\
similar item & 3.21 & 3 & 1.21 & 3.11 & 3 & 1.19 & * \\
\botrule
\end{tabular*}
\footnotetext{SD: Standard Deviation, *: $p<0.05$, **: $p<0.01$}
\end{table}

\subsection{Experiment 2}
The main result of Experiment 2 was the classification of buyers' information digestion. First, we calculated the importance of four page components (title, description, price, and image) in the evaluation experiments. Using conjoint analysis based on Quantification Theory Type 1, we determined the importance of page components for 60 combinations of participants and categories. The average importance values for each component were as follows: title 0.44, description 1.85, price 0.39, and image 0.73 (Table \ref{tab:result of quantification method of the first type}). The results indicated that participants prioritized information in the following order when understanding the product pages: description, image, title, and price. 

\begin{table}
\centering
\caption{Results of average part-worth utilities from Quantification Theory Type 1 and average ranges for each components.}\label{tab:result of quantification method of the first type}
\begin{tabular}{@{}llcccccc@{}}
\toprule
Attribute & Level & \multicolumn{3}{c}{Dummy} & Part-Worth Utility & Range  \\
\cmidrule{3-5}
 & & D1 & D2 & D3 & & \\
\midrule\midrule
title & short & 1 & 0 & - & 0.44 & 0.44 \\
        & long & 0 & 1 & - & 0.82 & \\ \midrule
price   & cheap & 1 & 0 & - & 0.67 &  0.38\\
       & expensive & 0 & 1 & - & 0.59 & \\ \midrule
product description & short & 1 & 0 & 0 & -0.42 & 1.85 \\
        & middle & 0 & 1 & 0 & 0.35 & \\
        & long & 0 & 0 & 1 & 1.34 & \\ \midrule
image     & 1 image & 1 & 0 & - & 0.27 & 0.73 \\
        & 3 images & 0 & 1 & - & 0.99 & \\
\bottomrule
\end{tabular}
\end{table}

\begin{figure}
\centering
\includegraphics[width=\textwidth]{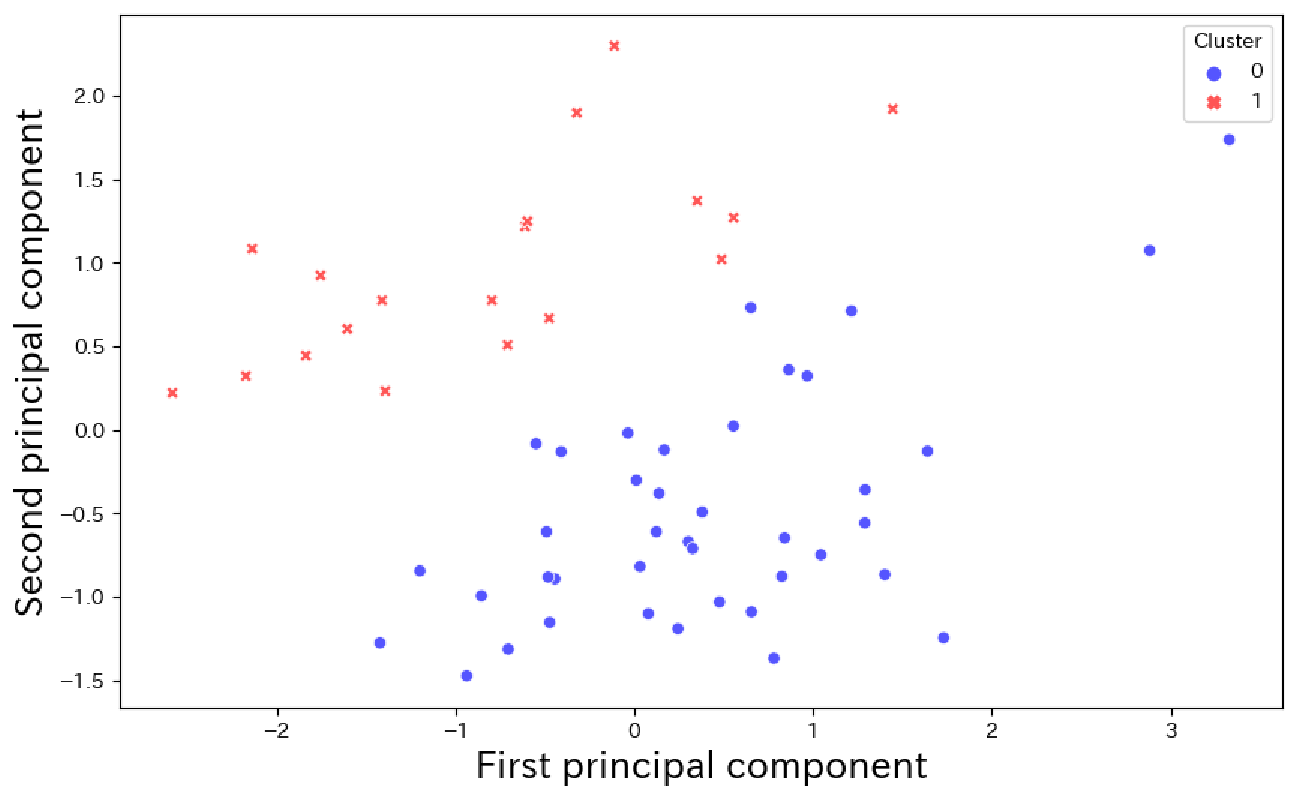}
\caption{The clustering results of the ranges for each attribute}\label{fig:clustering}
\end{figure}

Next, we applied X-means clustering (\cite{pelleg2000xmeans}) to the importance range data ($v_n$ = ($r_{title}$, $r_{price}$, $r_{description}$, $r_{figure}$), $n = 1, 2, \cdots, 60$) for 60 combinations of participants and categories. The analysis resulted in 40 instances in Cluster 0 and 20 instances in Cluster 1 (Figure \ref{fig:clustering}). X-means, an extension of K-means, determines the optimal number of clusters automatically using model selection criteria such as the Bayesian information criterion (BIC). In Figure \ref{fig:clustering}, for visualization, dimensionality reduction was performed using principal component analysis (PCA). We interpreted the first principal component as representing textual information and the second as representing information that could be quickly understood. Based on this interpretation, we labeled Cluster 0 as the group that carefully reads and digests information, and Cluster 1 as the group that processes information more quickly.

We analyzed the eye movement characteristics for each cluster using four indicators: TFF, FFD, TVD, and FC. The correlation coefficients between the importance range of each page component and the eye-tracking indicators were as follows for Cluster 0: TFF =  0.418, FFD = 0.594, TVD = 0.565, and FC = 0.521. In contrast, for Cluster 1, the coefficients were: TFF =  0.320, FFD = 0.422, TVD = 0.350, and FC = 0.33. Cluster 0 exhibited the highest correlation coefficients across all four indicators, We suggesting a stronger relationship between the importance range of page components and eye-tracking metrics. We statistically compared the correlation coefficients between the results of the conjoint analysis and these eye-tracking indicators across the clusters. The results indicated a significant difference $p$ $<$ 0.05 in the correlation coefficients between the range and TVD, as well as between the range and FC (Table \ref{tab:static_test_corr_by_cluster}).

\begin{table}
\centering
\caption{Statistical tests for correlation coefficients between ranges of part-worth utilities for each cluster and TFF, FFD, TVD, FC}\label{tab:static_test_corr_by_cluster}
\begin{tabular}{lcccc}
\toprule
 & TFF & FFD & TVD & FC \\
\midrule\midrule
z score & 0.812 & 1.681 & 1.978 & 2.147 \\
p value & n.s. & n.s. & * & *\\
\bottomrule
\end{tabular}
\footnotetext{*: $p<0.05$}
\end{table}

\section{Discussion}\label{discussion}
In Experiment 1, two major results were obtained. First, the importance of page components differs between buyers and sellers. Second, there was significant variation in the components that buyers considered important. These differences are likely due to the differing perspectives of buyers and sellers. Buyers' information needs and digestion processes are more context-dependent, varying with product category and familiarity with online transactions. However, sellers’ information provision strategies tend to be more consistent across product categories and experience levels. Sellers often provide similar information regardless of the product. These findings suggest that it is more effective for sellers to tailor the information provided to match the specific requirements of each product. Such an approach minimizes unnecessary information and mitigate information overload. 

In Experiment 2, we observed that buyers' information digestion could be divided into two clusters. One cluster required more time to digest the information, whereas the other processed the information more quickly. Regarding the eye-tracking results, TVD and FC are indicators that represent the extent of participants' attention to AOIs during the browsing. The high correlation coefficients in Cluster 0 between these indicators and the importance range suggest that Cluster 0 values the information by dedicating more viewing time to it. These eye-tracking results align with the PCA interpretations, strengthening the validity of the classification of information digestion.

Based on the results of the two experiments, we derived insights into three hypotheses related to the research question: Hypothesis 1: The difference in information digestion between buyers and sellers was confirmed through the analysis of product page components in Experiment 1, validating Hypothesis 1. Hypothesis 2: Experiments 1 and 2 showed that factors such as product type and transaction history influence information digestion processes, supporting Hypothesis 2. Hypothesis 3: Through the analysis in Experiment 2, we successfully identified the process of buyers' information digestion and clustered their tendencies, validating Hypothesis 3. The validation of these hypotheses signifies the critical role of the information processing in C2C markets and underscores the importance of tailoring information to meet the diverse needs of buyers. 

\section{Conclusion}\label{conclusion}
This study explored the differences in information digestion between buyers and sellers in the online C2C market through two experiments: a questionnaire survey and an eye-tracking experiment. By conducting the questionnaire survey, we identified differences in the importance assigned to product page components by buyers and sellers. We also revealed that buyers' information digestion processes were influenced by product categories and transaction experience. Moreover, by evaluating practical product pages and analyzing eye-tracking measurements, we classified buyers' information digestion processes into distinct patterns. The findings from these experiments provided insights into how buyers and sellers processed and evaluated product information differently, facilitating more effective information communication and bridging the gaps between them.

Future research could build on this work by exploring the impact of product page layouts on information digestion. Increasing the sample size in future studies would allow for more robust analyses and enhance the generalizability of the results. In addition, investigating the long-term impacts of improved information digestion on trust and transaction success in C2C markets could yield deeper insights into its benefits.

This study contributed to the understanding of information digestion in online C2C markets and added to the body of knowledge on trust-building mechanisms. It also offered practical insights for platform designers and market participants. By tailoring product information to better align with buyers' needs and preferences, C2C platforms could improve the user experience, foster trust, and drive market growth.

\section*{Acknowledgements}
This study was supported by the joint research on Value Exchange Engineering conducted with Mercari R4D and RIISE, as well as by JSPS KAKENHI Grant Number JP20H02384. 

\section*{Informed Consent}
This study was conducted in accordance with the relevant ethical guidelines, following review and approval the Ethics Committee of the Graduate School of Engineering at the University of Tokyo (examination number: 23-122; approval number: KE23-128). All participants were provided with written explanations regarding the purpose and methodology of the study, its risks and benefits, privacy protection measures, and the voluntary nature of participation. Written informed consent was obtained from each participant after confirming their full understanding.

\end{document}